\documentclass[10pt]{article}
\usepackage{amsmath}
\usepackage{graphicx,subfigure,float}
\usepackage[varg]{txfonts}

\newcommand{\vect}[1]{\ensuremath{\boldsymbol{#1}}}
\newcommand{\tens}[1]{\ensuremath{\boldsymbol{\mathsf{#1}}}}

\newcommand{\DD}{\ensuremath{\mathrm{D}}}

\title{The Numerical Simulation of Turbulence}

\author{W. Schmidt
    \medskip\\
    {\normalsize Institut f\"{u}r Theoretische Physik und Astrophysik, Universit\"{a}t W\"{u}rzburg}
}

\date{\normalsize Proceedings contribution to
    \medskip\\
	\textbf{Structure formation in the Universe: Chamonix 2007}}

\begin{document}

\maketitle

Turbulence is a remarkable subject in physics. The underlying
equations, which are in their simplest formulation the Euler
equations, were published 250 years ago \cite{Euler57}. Yet a
theoretical grasp of the phenomenology emerging from these equations
had not been achieved before the mid-twentieth century, when
Heisenberg~\cite{Heisenberg} and Kolmogorov~\cite{Kolmog41} obtain
first analytical results. Eventually, it took the capabilities of
modern supercomputers to obtain a full appreciation of the complexity
that is inherent to the Euler equations. Astrophysics is now at the
very frontier of numerical turbulence modelling. Among the additional
ingredients for making turbulence in astrophysics even more complex 
are supersonic flow, self-gravity, magnetic fields and radiation
transport. In contrast, terrestrial turbulence is mostly incompressible or
only weakly compressive. External gravity is, of course, an issue in the
computation of atmospheric processes on Earth. Self-gravity, however,
is only encountered on large, astrophysical scales. The dynamics of
turbulent plasma has met vivid attention in research related to
nuclear fusion reactors but, otherwise, is not encountered under
terrestrial conditions.

In this Chapter, I give an overview of the various approaches toward
the numerical modelling of turbulence, particularly, in the
interstellar medium (ISM). The discussion is placed in a physical
context, i.~e. computational problems are motivate from basic physical
considerations. Presenting selected examples for solutions to these
problems, I introduce the basic ideas of the most commonly used
numerical methods. For detailed methodological accounts, the reader is
invited to follow the references. Some important results and
astrophysical implications are briefly outlined. Since turbulence, in
a strict sense, is genuinely three-dimensional, I almost exclusively
consider three-dimensional simulations. Furthermore, turbulence is a
multi-scale phenomenon and, in order to capture its properties
correctly, sufficient numerical resolution is essential. This is why
newer higher-resolution computations are generally preferred as
examples (as regards self-gravitating turbulence, however,
high-resolution simulations are appearing just now and are not
included here).

To begin with, I briefly discuss the equations which are numerically
solved in astrophysics and related theoretical aspects. The following
Sections deal with supersonic turbulence, self-gravitating turbulence
and magnetohydrodynamic turbulence. Naturally, there are
intersections. Representative examples for numerical simulations are placed
depending on their main objectives. Finally, I give a resume of what
has been achieved and which challenges are likely to be met.

\section{Fundamentals}
\label{sc:fundament}

In the following, we consider self-gravitating inviscid fluid subject
to the action of external force fields and, possibly, magnetic
fields. The evolution of the mass density $\rho$, the velocity $v$ and
the specific energy $e$ of the fluid is given by the compressible
Euler equations:
\begin{align}
  \label{eq:mass}
  &\frac{\DD}{\DD t}\rho = -\rho\vect{\nabla}\vect{v}, \\
  \label{eq:vel}
  \rho &\frac{\DD}{\DD t}\vect{v} =
  -\vect{\nabla}P + \rho(\vect{f} + \vect{g}), \\
  \label{eq:energy}
  \rho &\frac{\DD}{\DD t}e + \vect{\nabla}\cdot\vect{v}
P =
  \Gamma -\Lambda + \rho\vect{v}\cdot(\vect{f} + \vect{g}),
\end{align}
where the Lagrangian time derivative is defined by
\begin{equation}
    \frac{\DD}{\DD t} =
    \frac{\partial}{\partial t} + \vect{v}\cdot\vect{\nabla}.
\end{equation}
The total energy per unit mass is given by
\begin{equation}
    e = \frac{1}{2}v^{2} + \frac{P}{(\gamma-1)\rho},
\end{equation}
where $\gamma$ is the adiabatic exponent and the pressure $P$ is
related to the mass density and the temperature $T$ via the ideal
gas law:
\begin{equation}
    \label{eq:ideal_eos}
    P=\frac{\rho k_{\mathrm{B}}T}{\mu m_{\mathrm{H}}}.
\end{equation}
The constants $k_{\mathrm{B}}$, $\mu$ and $m_{\mathrm{H}}$ denote,
respectively, the Boltzmann constant, the mean molecular weight and
the mass of the hydrogen atom. The energy budget can be altered by
heating ($\Gamma$) and cooling ($\Lambda$), by non-gravitational
forces ($\rho f$) as well as gravity ($\rho g$). Generally, heating
and cooling processes are important for the dynamics of the
interstellar medium (Sections~\ref{sc:grav} \& \ref{sc:mhd}).

Non-gravitational forces supplying energy to the flow can be
mechanical or magnetic. An example of a mechanical force would be the
random driving force that is commonly used in turbulence
simulations. This is an external force, i.~e., it is independent of the
dynamics of the fluid (Section~\ref{sc:supersonic}). Quite the
opposite holds for conducting fluids in the presence of magnetic
fields. In the case of ideal magnetohydrodynamics, the fluid is
dragged by the Lorentz force $\rho\vect{f}=\vect{J}\times\vect{B}$,
where the current $\vect{J}=\vect{\nabla}\times\vect{B}$. The magnetic
field $\vect{B}$, in turn, depends on the flow via Faraday's
law,
\begin{equation}
  \frac{\partial\vect{B}}{\partial t} =
  \vect{\nabla}\times(\vect{v}\times\vect{B}).
\end{equation}
The interaction between turbulent flow and magnetic fields alters
properties of turbulence such as the scaling behaviour (Section~\ref{sc:mhd}).

The gravitational acceleration $\vect{g}=-\nabla\cdot\phi$ arises, in part,
from the self-gravity of the fluid and, depending on the boundary conditions, the
gravitational field of external sources. In the case of periodic boundary conditions, gravity is
solely produced by density fluctuations with respect to the global
mean (Section~\ref{sc:grav}). In this case, the gravitational
potential is determined by the Poisson equation
\begin{equation}
  \label{eq:poisson}
  \nabla^{2}\phi = 4\pi G(\rho-\langle\rho\rangle),
\end{equation}
where $G$ is Newton's constant. The mean mass density
$\langle\rho\rangle=\rho_{0}$ is a constant because of mass
conservation. 

Letting the curl operator $\vect{\nabla}\times$ act upon
equation~(\ref{eq:vel}), an evolutionary equation for the vorticity of the flow,
$\omega=\vect{\nabla}\times\vect{v}$, is obtained:
\begin{equation}
  \label{eq:vort}
  \frac{\DD}{\DD t}\vect{\omega} =
  \tens{S}\cdot\vect{\omega} - d\vect{\omega}
  + \frac{1}{\rho^{2}}\vect{\nabla}\rho\times\vect{\nabla}P +
  \vect{\nabla}\times\vect{f}.
\end{equation}
The rate-of-strain tensor $\tens{S}$ is the symmetrised velocity
gradient and the divergence $d=\vect{\nabla}\cdot\vect{v}$. Apart
from the stirring of fluid due to rotational force components,
vorticity can be generated by the baroclinic term
$\vect{\nabla}\rho\times\vect{\nabla}P$, if the gradient of the mass density and the
pressure gradient are not aligned. With the ideal gas
equation~\ref{eq:ideal_eos}, it follows that baroclinic vorticity
generation occurs in non-isothermal gas. The term
$\tens{S}\cdot\vect{\omega} - d\vect{\omega}$
accounts for the stretching of vortices caused by strain in addition to the
expansion or contraction of vortices due to the compressibility of the fluid.

The time evolution of the divergence $d$ is given by
\begin{equation}
  \begin{split}
  \label{eq:div}
  \frac{\DD}{\DD t}d
  = & \frac{1}{2}\left(\omega^{2}-|S|^{2}\right) -
      \frac{1}{\rho}\nabla^{2}P \\
    & + \frac{1}{\rho^{2}}\vect{\nabla}\rho\cdot\vect{\nabla}P -
        4\pi G(\rho-\rho_{0}) +
        \vect{\nabla}\cdot\vect{f}.
  \end{split}
\end{equation}
This equation also follows from the conservation law~(\ref{eq:vel}) by
applying the operator $\vect{\nabla}\cdot$ and substituting the
Poisson equation for the gravitational potential~(\ref{eq:poisson}).
The norm of the rate of strain is defined by
$|S|=(2S_{ij}S_{ij})^{1/2}$. From the first term, one can see that
vorticity contributes to positive divergence, while strain is associated
with negative divergence, i.~e. converging flow. In the incompressible
case, the first term on the right hand side of equation~(\ref{eq:div}) is
cancelled by the Laplacian of the pressure divided by the
density. Whereas vorticity cannot be generated by a gradient field
such as gravity, an increasing rate of convergence will be caused by
gravity, if the term $4\pi G(\rho-\rho_{0})$ dominates. In this case, gravitational collapse
of the gas ensues.

The numerical simulation of astrophysical turbulence has to account for all
fluid dynamical processes which become manifest in the evolutionary
equations of the vorticity~(\ref{eq:vort}) and the
divergence~(\ref{eq:div}). In many terrestrial applications, on the
other hand, incompressible turbulence without gravity and magnetic
fields is considered. Then equations~(\ref{eq:vort})
and~(\ref{eq:div}) reduce to
\begin{align}
  & \frac{\DD}{\DD t}\vect{\omega} =
    \tens{S}\cdot\vect{\omega} +
    \vect{\nabla}\times\vect{f}, \\
  & \frac{\DD}{\DD t}d = 0.
\end{align}
The only possibility of generating vorticity in this case is to stir
the fluid by \emph{solenoidal} (rotational) forces. Non-linear
turbulent interactions stretch and fold the large-scale eddies
produced by stirring into thin vortex filaments. This is the very
essence of turbulent fluid dynamics. Here, the question
arises whether the generation of turbulence in the ISM
resembles the scheme of stirring, stretching and folding at all. 
Interestingly, this presumption underlies the application
of solenoidal driving forces in most numerical simulations.

Incompressible fluid dynamics serves as an
important reference case for which scaling properties of isotropic
turbulence in the ensemble average can be derived analytically. 
Kolmogorov~\cite{Kolmog41} showed that the root mean square
velocity fluctuations $v'(\ell)$ at a certain length scale $\ell$
obey the so called $2/3$-law,
\begin{equation}
    v'(\ell)^{2}\propto\ell^{2/3},
\end{equation}
provided that $\ell$ is small compared to the integral length scales
$L$ at which energy is supplied to the flow. The 2/3-law implies that
the rate of kinetic energy dissipation $\epsilon$ is asymptotically
constant, as $\epsilon\sim v'(\ell)^{3}/\ell\simeq\mathrm{const.}$ in
the limit $\ell/L\ll 1$. This result is known as the law of positive
energy dissipation \cite{Frisch}. Both the $2/3$ law and the law of
positive energy dissipation appear to be robust properties of
turbulence as long as the flow becomes isotropic and nearly
incompressible towards small length scales.  This is not always the
case. Magnetic fields, for instance, introduce small-scale
anisotropy. Moreover, there is no consensus yet to what extent the
scaling properties of incompressible turbulence carry over to
supersonic flow.

\section{Supersonic turbulence}
\label{sc:supersonic}

Turbulence is called supersonic, if the root mean square (RMS) Mach
number
$\mathcal{M}_{\mathrm{rms}}=\langle(v/c_{\mathrm{s}})^{2}\rangle^{1/2}
$ exceeds unity. Kinetic energy dissipation in supersonic turbulence
caused by shock fronts propagating through the fluid dominates viscous
dissipation. Since the resulting dissipative heating rises the speed
of sound to a value comparable to the typical flow velocity within a
dynamical time, supersonic turbulence cannot be maintained, if the
excess heat is not efficiently removed from the fluid.  As an
idealisation, heating is exactly balanced by cooling such that the
temperature remains constant. In this case, continuously driven
turbulent flow settles into a steady state in which the RMS Mach
number is asymptotically constant. In isothermal gas, solenoidal
forcing is required to produce turbulence because the density gradient
is aligned with the pressure gradient and, consequently,
$\vect{\nabla}\rho\times\vect{\nabla}P=0$.  In this case, it can be
seen from equation~(\ref{eq:vort}) that non-zero vorticity either must
be imposed as initial condition or it is generated by the curl of the
force field. As a consequence of the constant speed of sound, the
properties of driven supersonic turbulence in isothermal gas depend on
a single parameter only, namely, the RMS Mach number (provided that
the forcing is purely solenoidal \cite{SchmFeder07}).

A a lot of effort has gone into the numerical simulation of supersonic
flow by means of finite-volume schemes of higher order, in particular,
the piecewise parabolic method (PPM) \cite{ColWood84}. The
applicability of PPM to turbulence was tested in simulations of
adiabatic turbulence \cite{SytPort00} and nearly isothermal transonic
turbulence.  An example for the latter is the driven turbulence
simulation by Porter \& Pouquet \cite{PortPou02}. They utilised an
explicit cooling function and, thereby, maintained a root mean square
Mach number $\approx 1.0$.  In \cite{SchmHille06}, the equation of
state of degenerate electron gas was applied in simulations of forced
transonic turbulence with PPM. Degeneracy entails an extremely large
heat capacity. Since the isothermal case corresponds to the limit of
infinite heat capacity, the gas in these simulations is effectively
isothermal. One of the conclusions was that the numerical dissipation
of PPM acts as an implicit subgrid scale model in the weakly
compressible regime (at small scales).

A new approach is the application of adaptive mesh refinement (AMR) in
turbulence simulations \cite{KritNor06}. AMR is based on the
finite-volume approach with a hierarchy of grid patches of different
resolution. The basic idea is to represent relatively smooth flow
regions on coarse grids only, whereas steep gradients of the velocity,
density or any other field are computed with higher accuracy by
dynamically inserting grids of higher resolution
\cite{BerOli84,BerCol89}. In astrophysics, AMR has been successfully
used for the treatment of strong shocks or gravitational collapse in a
variety of problems \cite{SheaBry04}. On the other hand, the
simulation of developed turbulence with AMR has been regarded as
infeasible, because, in the picture of the turbulent cascade of
eddies, small-scale features of the flow would be space-filling
\cite{Frisch}. For a particular flow realization, however, turbulent
eddies occupy an ever decreasing fraction of the flow domain toward
smaller length scales at any given instant of time. This follows from
the intermittency of turbulence \cite{Frisch}.

Picking up the intermittent picture of turbulence, Kritsuk et al.\
proposed that AMR would offer a computational advantage compared to
static grids of fixed resolution \cite{KritNor06}. This can be
motivated from the Landau estimate of the number of degrees of
freedom \cite{LandLif},
\begin{equation}
    N\sim\mathrm{Re}^{9/4},
\end{equation}
where $\mathrm{Re}$ is the Reynolds number of the flow. Present day
supercomputers manage $N\sim10^{10}$, which allows for $\mathrm{Re}
\sim 10^{4}$. Note that fully developed turbulence is known
to set in at Reynolds numbers of a few thousand. On the other hand, invoking
intermittency models of turbulence such as the $\beta$-model
\cite{Frisch}, we have
\begin{equation}
    N\sim\mathrm{Re}^{3D/(D+1)},
\end{equation}
where $D$ is interpreted as the fractal dimension of dissipative
structures. Assuming a value of $D\approx 2$ as in the Burger model of
supersonic turbulence, it follows that $N\sim 10^{8}$. This figure is
smaller than the Landau estimate (corresponding to a fixed-resolution
grid) by two orders of magnitude. Of course, there is substantial
computational overhead with AMR in comparison to static grids. Apart
form that, more sophisticated intermittency models and numerical
studies suggest that $D>2$ in supersonic turbulence
\cite{Boldyrev02,BoldyNord02}.  Nevertheless, AMR is expected to pay
off, if several levels of refinement are used. In particular, this
applies to scenarios including gravitational collapse (see the next Section).

\begin{figure}[t]
    \centering
    \includegraphics[width=0.8\linewidth]{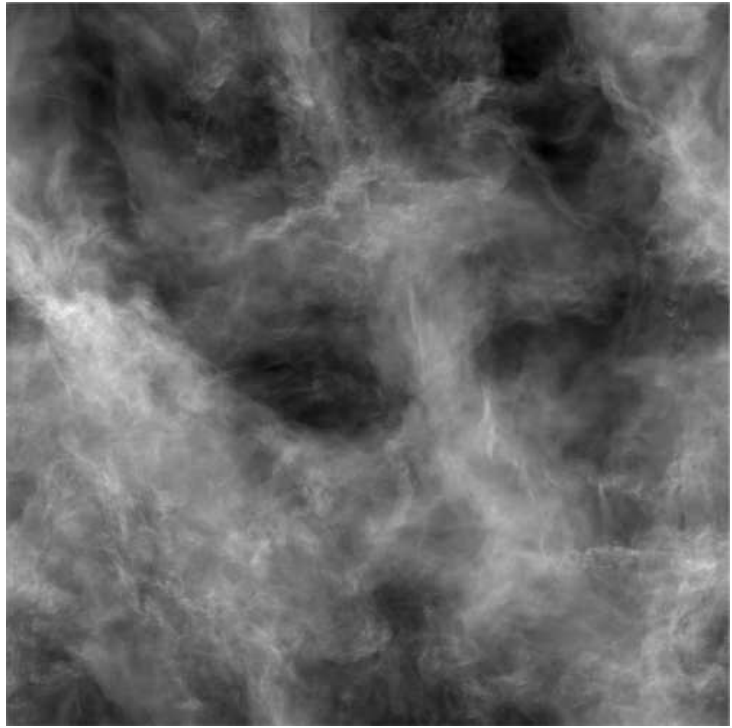}
    \caption{Projected mass density in an AMR simulation of supersonic 
	turbulence with effective resolution $2048^{3}$ by Kritsuk et al.\, 2007, using Enzo
	\cite{KritNor07}. Bright regions contain gas of high density, while
	under-dense gas appears dark.}
    \label{fg:amr_turb}
\end{figure}

In \cite{KritNor07}, the effective resolution (i.~e. the resolution
corresponding to the most refined grids) of an AMR simulation of
supersonic turbulence was raised to the yet unprecedented value of
$2048^{3}$. Nearly isothermal gas was maintained by setting the
adiabatic exponent $\gamma$ to a value that differs only by a small
fraction from unity. This means that the internal energy
$E_{\mathrm{int}}=P/(\gamma-1)\rho$ is artificially increased by a huge factor such
that $E_{\mathrm{int}}\gg \frac{1}{2}\rho v^{2}$ and, hence, it takes
many dynamic time scales to heat the gas significantly by kinetic
energy dissipation. In combination with a $1024^{3}$ static grid simulation,
a great wealth of results was obtained from the analysis of the
numerical data. Figure~\ref{fg:amr_turb} shows the projected
mass density in logarithmic scaling for one snapshot of the AMR
simulation. One can see voids in between high-density regions which display
intricate turbulent structure. This is a tell-tale sign of the
pronounced intermittency of supersonic turbulence.

Further indications of intermittent properties come from the probability
density function of the time-averaged mass density, which follows
very closely a log-normal distribution \cite{PadNord02}, and
the scaling exponents of the velocity structure functions. The
definition of structure functions is based on spatial
correlations of the velocity field:
\begin{equation}
    S_{p}(\ell)=
    \langle|\vect{v}(\vect{x}+\vect{\ell})-\vect{v}(\vect{x})|\rangle^{p}.
\end{equation}
The square brackets denote the average over the whole domain of the
flow. Structure functions probe the correlations of the velocity
field at different spatial positions depending on the separation
$\ell$. It is both a theoretical prediction and an experimentally
well established fact that the structure functions of isotropic turbulence
obey power laws
\begin{equation}
    S_{p}(\ell)\propto\ell^{\zeta_{p}}.
\end{equation}
The exponents $\zeta_{p}$ are called the scaling exponents. For
incompressible turbulence, Kolmogorov found $\zeta_{p}=p/3$. In the
case of the second-order structure functions ($p$=2) this result
corresponds to the $2/3$-law for the velocity fluctuations mentioned
in Section~\ref{sc:fundament}.  Calculating the scaling exponents form
their numerical data, Kritsuk et al.\ were able to demonstrate
significant deviations form the Kolmogorov relation in accordance with
predictions of the intermittency model proposed by Boldyrev et al.\
\cite{BoldyNord02,KritNor07}, which is a generalisation of the
She-L\'{e}v\^{e}que model for incompressible turbulence
\cite{SheLeveq94}. In particular, it was found that $\zeta_{2}\approx
1.0$, whereas $\zeta_{2}=2/3$ in the Kolmogorov theory. Intriguingly,
the analysis also revealed that the scaling exponents applicable to
incompressible turbulence are obtained, if the statistics is computed
for the density-weighted variable $\rho^{1/3}\vect{v}$ in place of
$\vect{v}$, a result that is not fully understood yet but might bear
important implications on the nature of supersonic turbulence.

\section{Self-gravitating turbulence}
\label{sc:grav}

According to the linear stability analysis of the Euler equations by 
Jeans \cite{Jeans02}, a perturbation in isothermal gas of temperature 
$T$ and uniform density $\rho_{0}$ becomes unstable against 
gravitational collapse, if its size exceeds the Jeans length \begin
{equation}
	\lambda_\mathrm{J} = c_\mathrm{s}\sqrt{\frac{\pi}{G\rho_{0}}},
\end{equation}
where $c_\mathrm{s}\propto T^{1/2}$ is the isothermal speed of sound. 
Beginning with Chandrasekhar's proposition to substitute $c_\mathrm
{s}^{2}$ by 
\begin{equation}
	\label{eq:chandra}
	c_\mathrm{eff}^{2} =
	c_\mathrm{s}^{2}+\frac{1}{3}\langle v^{2}\rangle
\end{equation}
in the above expression for the Jeans length, several
attempts have been made to extend the Jeans stability analysis to 
turbulent gas of RMS velocity $\langle v^{2}\rangle^{1/2}$ 
\cite{Chandra51,BonaFal87,BonaPer92}.
Equation~(\ref{eq:chandra}) implies that the effective gas pressure is given by the sum of 
the thermal pressure $P$ and the turbulence pressure $P_
\mathrm{turb}=\frac{2}{3}E_{\mathrm{kin}}$, where $E_{\mathrm{kin}}$
is the mean kinetic energy density.

However, even the most advanced analysis put forward so far \cite{BonaPer92},
suffers from severe constraints. A perturbation analysis can be carried out for a 
statistically stationary equilibrium state only. Thus, it has to be 
assumed that the free fall time scale $T_{\mathrm{ff}}\sim(G\rho_{0})^{-1/2}$
is much greater than dynamical time scale of turbulent flow. 
From the divergence equation~(\ref{eq:div}), one can see that this 
assumption implies that the self-gravity term remains small compared 
the other terms at all times. Allowing for gravitational
collapse, however, self-gravity eventually dominates and
\begin{equation}
	\frac{\DD}{\DD t}d \sim -G\rho.
\end{equation}
A non-perturbative theory of the regime in which the free-fall time 
scale and the dynamical time scale of turbulence are comparable was 
suggested by Biglari and Diamond \cite{BigDia89}. Combining scaling 
relations from the $\beta$-model of turbulence \cite{Frisch} with the 
assumption of energy equipartition between gravity and turbulence at 
all scales, they derived an intermittent hierarchical cloud model of self-gravitating turbulence. 

At present, however, there is no general theory that would fully
encompass the highly non-linear and non-stationary interplay between
gravity and turbulence in the interstellar medium. Numerical
simulations, on the other hand, can aid to the understanding
self-gravitating turbulence, although one has to resort to artificial
mechanisms of producing turbulence or imposing more or less arbitrary
initial conditions. One technique is to apply a driving force as
mentioned in the previous Section. Alternatively, some random initial
velocity field might be assumed as initial condition. This results in
decaying turbulence. Yet another option is to consider gas in an
unstable state and apply small perturbations. The most prominent
example is the thermal instability \cite{KritNor02,ElmeSca04}.

\begin{figure}[t]
    \centering
    \includegraphics[width=0.67\linewidth]{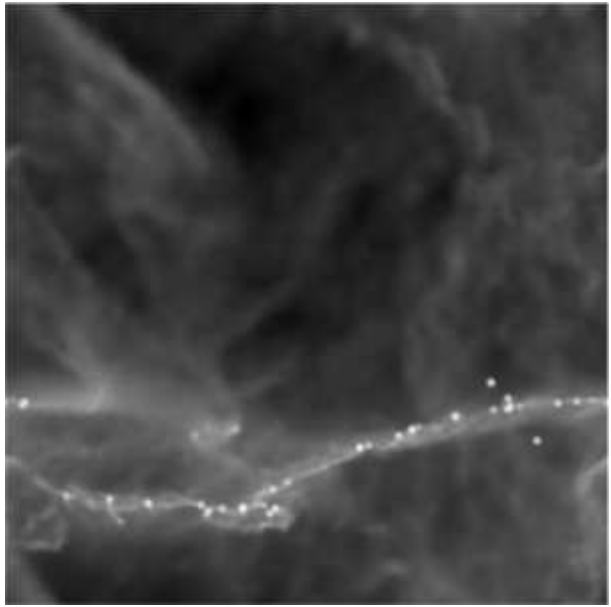}
    \caption{Filamentary structure with sink particles in an SPH 
simulation of self-gravitating turbulence (courtesy of R. Klessen). In 
this plot, over-dense regions appear bright.}
    \label{fg:filament_sph}
\end{figure}

\begin{figure}[t]
    \centering
    \includegraphics[width=0.67\linewidth]{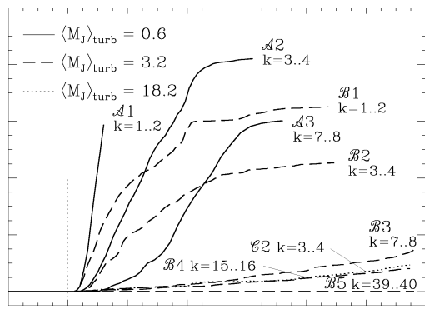}
    \vspace{0.07\linewidth}
    \caption{Total mass fraction captured by sink particles as a 
function of time in SPH simulations with different Mach numbers and 
varying driving scale of turbulence (Klessen et al., 2000 \cite
{KlessHeitsch00}).}
    \label{fg:star_prod}
\end{figure}

Numerical simulations of self-gravitating driven or decaying
supersonic turbulence were initially performed with smoothed particle
hydrodynamics (SPH) because of its ability of tackle variations of the
mass density over many orders of magnitude \cite{Mona92}. SPH makes
use of the Lagrangian framework of fluid mechanics and represents
fluid parcels by particles.  Examples for the application of SPH to
self-gravitating turbulence are the simulations by Klessen et al.\
\cite {Kless00,KlessHeitsch00,Kless01}. In these simulations, the
influence of self-gravity on probability density functions
\cite{Kless00} and the fragmentation properties of the gas was
investigated both for driven and for decaying supersonic turbulence
\cite{KlessHeitsch00,Kless01}. Sink particles were introduced to
capture collapsing regions beyond a certain density threshold in order
to prevent the mass density from growing indefinitely.
Figure~\ref{fg:filament_sph} illustrates that sink particles are
mostly formed within filamentary structures (which are also seen in
cosmological simulations). Varying the Mach number, it was found that
higher turbulence intensity slows down the rate at which sink
particles are produced considerably (Figure~\ref{fg:star_prod}). The
major conclusion drawn by Klessen et al.\ was that supersonic
turbulence inhibits gravitational collapse globally, as one would
suspect from equation~(\ref {eq:chandra}). The strong gas compression
caused by shocks, on the other hand, can trigger gravitational
collapse locally and initiate the formation of stars.

Jappsen et al.\ \cite{JappKless05} studied the mass spectrum produced
by turbulent fragmentation in non-isothermal gas. They defined the
state variables by piecewise polytropic relations,
i.~e. $P\propto\rho^{\gamma_{\mathrm{eff}}}$ for a certain range of
densities. If $\gamma_{\mathrm{eff}}<1$, the gas cools with increasing
density. This effect enhances the compressibility of the gas and
therefore supports gravitational collapse (negative
$\vect{\nabla}\rho\cdot\vect{\nabla}T$ contribution adds to the
gravity term in the divergence equation~\ref{eq:div}) . At some
threshold density, however, the cooling regime ceases and
$\gamma_{\mathrm{eff}}$ becomes greater than unity. From various
simulations, it was found that the transition from the cooling regime
to isothermal or nearly isothermal gas sets a characteristic mass
scale of turbulent fragmentation. This mass scale was interpreted to
correspond to the peak of the observed initial mass
function. V\`{a}zquez-Semadeni et al.\ \cite{VazGom07} included
explicit heating and cooling in SPH simulations of colliding gas
streams. They made use of a model for the cooling function $\Lambda$
in the energy equation~(\ref{eq:energy}) that was proposed by Koyama
\& Inutsuka \cite{KoHiro02}. Due to the thermal instability
(compression causes the gas to cool), the gas at the collision
interface undergoes gravitational collapse and becomes increasingly
turbulent. This is interpreted as possible mechanism of molecular
cloud formation. Remarkably, approximate equipartition of
gravitational and kinetic energy was found, although clearly no state
of virial equilibrium was approached by the collapsing gas. Further
applications of SPH addressing problems such as the formation of brown
dwarfs, binary star systems and stellar clusters via turbulent
fragmentation as well as the origin of the initial mass function were
presented by Bate et al. \cite{BateBon02a,BateBon02b,BateBon05} and
Bonnel et al.  \cite{BonBate03}.

Despite the numerous important contributions to the understanding of
the role of turbulence to star formation, the adequacy of treating
turbulence with SPH has been a matter of debate. On the one hand,
numerical studies of self-gravitating systems indicate basic agreement
between AMR and SPH \cite{CommHenne07}. However, comparisons with
grid-base codes suggest that SPH dissipates small-scale velocity fluctuations
significantly stronger \cite{PadNord07,Kitsion07}. In
particular, it was demonstrated that the steeper power spectra
obtained from SPH simulations are accompanied by mass distributions
which are not consistent with observations. Apart from that, it
appears to be rather difficult to accommodate magnetohydrodynamics in
the SPH formalism. Although there are ongoing attempts to overcome
this shortcoming \cite{MarHow03}, MHD has been treated with great
success using finite-volume discretisation.

\section{Magnetohydrodynamic turbulence}
\label{sc:mhd}

In purely hydrodynamical turbulence, there is no preferred spatial
direction for small-scale velocity fluctuation. The randomisation of
the flow due to non-linear turbulent interactions produces
statistical isotropy towards small scales which can be interpreted
as a symmetry of the ensemble average \cite{Frisch}. This symmetry
is broken by a magnetic field $\vect{B}$ in turbulent conducting
fluid, if the ratio of the thermal to the magnetic pressure,
\begin{equation}
    \beta=
    \frac{P}{P_{\mathrm{M}}}=
    \frac{8\pi P}{B^{2}},
\end{equation}
is comparable to unity or less. If the curl of the Lorentz force,
$\vect{\nabla}\times(\vect{J}\times\vect{B})$, in the vorticity
equation~(\ref{eq:vort}) is sufficiently strong, then the vorticity
will mainly grow in the direction of $\vect{B}$. In developed
magnetohydrodynamic (MHD) turbulence, this effect causes eddies to
shrink perpendicular to the field lines. As a further implication, the
anisotropy of MHD turbulence increases towards smaller length scales
\cite{Biskamp}.

\begin{figure}[thb]
    \centering
    \includegraphics[width=0.67\linewidth]{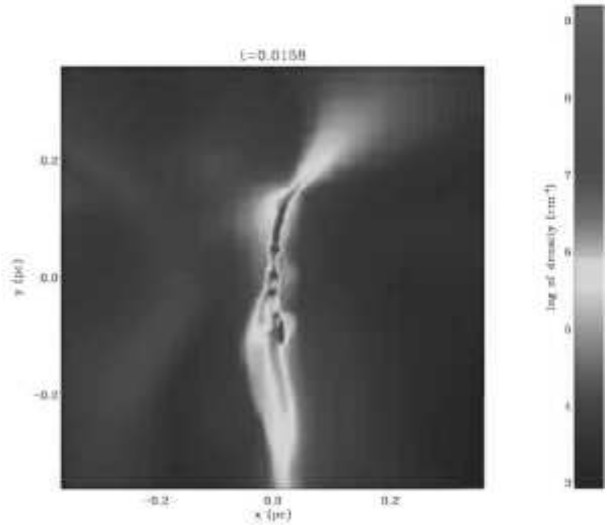}
    \caption{Simulation of a collapsing gas cloud of $1000$ solar masses  with Ramses (courtesy of P. Hennebelle). The magnetic energy of the clouds is about the same as its kinetic energy.}
    \label{fg:mhd_cloud}
\end{figure}

\begin{figure}[thb]
  \begin{center}
  	\includegraphics[width=0.8\linewidth]{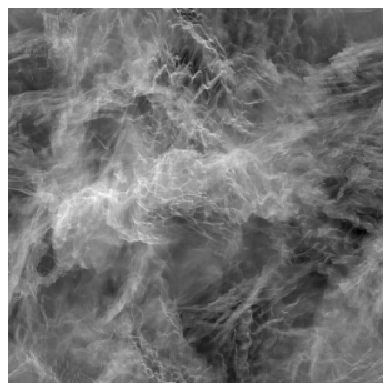}
	\caption{Projected mass density in a simulation of hydrodynamic turbulence with the Stagger code (Padoan et al., 2007 \cite{PadNord07}). For comparison with MHD turbulence, see Figure~\ref{fg:mhd}}.
    \label{fg:hd}
  \end{center}
\end{figure}

\begin{figure}[thb]
  \begin{center}
	\includegraphics[width=0.8\linewidth]{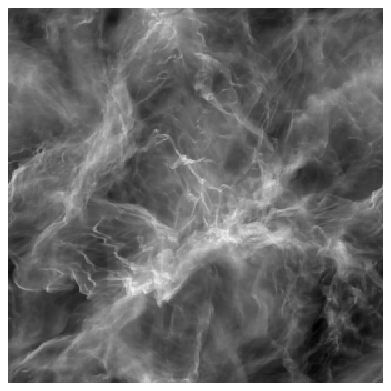}
	\caption{Projected mass density in a simulation of magnetohydrodynamic turbulence with the Stagger code (Padoan et al., 2007 \cite{PadNord07}). For comparison with purely hydrodynamic turbulence, see Figure~\ref{fg:hd}}.
    \label{fg:mhd}
  \end{center}
\end{figure}

The numerical simulation of MHD turbulence is rather
challenging. Godunov-based schemes of higher order such as PPM become
very complex upon including MHD. For this reason, simpler schemes have
been adopted for MHD. Examples are the Zeus code \cite{StoneNor92},
the Stagger code \cite{NordGals95}, and the Ramses code which also
features AMR \cite{Teyssier02,ForHenne06}. As an illustration,
Figure~\ref{fg:mhd_cloud} shows a simulation of the gravitational
collapse of a magnetised cloud with up to 6 levels of refinement. One
of the major problems that needs to be addressed when solving the MHD
equations numerically is to keep the magnetic field divergence-free,
i.~e.  $\vect{\nabla}\cdot\vect{B}=0$. In Ramses and other codes, the
constrained transport algorithm is employed to ensure this
constraint \cite{EvansHaw88}.

A comparison of the properties of driven hydrodynamic (HD) and MHD
turbulence in high-resolution simulations was presented by Padoan et
al.\ \cite{PadNord07}. Whereas the energy spectrum functions,
irrespective of magnetic fields, were found to follow closely the
power law $k^{-2}$ in the inertial subrange, the fragmentation
properties of the gas appear to be markedly different for HD and MHD
turbulence, respectively. In plots of the projected density fields for
both cases (Figures~\ref{fg:hd} and~\ref{fg:mhd}), one can see that
the gas is more concentrated in pronounced filaments under the action
of magnetic fields. Utilising a clump-find algorithm, it was
demonstrated that a significantly steeper mass spectrum of dense cores
(defined by a density threshold based on the Bonnor-Ebert mass)
resulted from purely hydrodynamical turbulence. Furthermore, they were
able to reproduce the Chabrier initial mass function \cite{Chab03} in
the MHD case, although this result must be considered with care given
the ambiguity of calculating mass spectra.

Glover and Mac Low \cite{GlovLow07} performed simulations of
self-gravitating MHD turbulence including various thermal and chemical
processes. They were able to show that the fast formation of molecular
hydrogen within a few Myrs, which is typically observed in molecular
clouds, can be explained as an effect of turbulence. Hennebelle \&
Audit \cite{HenneAud07} investigated thermally bistable MHD turbulence
and found mass spectra similar to those inferred from CO
observations of molecular clouds. Although first computed in only 2 dimensions, the setup
was generalised to include self-gravity and computed in 3 dimensions
as well. The ansatz by Hennebelle et al.\ is different from
simulations of isothermal MHD turbulence as it does not rely on a
driving force or an initial turbulent velocity field.

\begin{figure}[thb]
  \begin{center}
    \mbox{\subfigure{\includegraphics[width=0.48\linewidth]{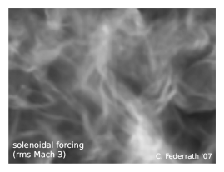}}\quad
    	\subfigure{\includegraphics[width=0.48\linewidth]{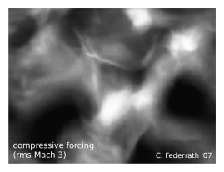}} }
	\caption{Projected mass density in $128^{3}$ simulations of turbulence driven by a solenoidal and a compressive force field, respectively (courtesy of C. Federrath).}
    \label{fg:forcing}
  \end{center}
\end{figure}

\section{Perspectives}

Up to now, most approaches to the numerical simulation of
astrophysical turbulence have focused on particular aspects, be it a
thorough understanding of supersonic turbulence, the study of the
dynamics of self-gravitating turbulent gas or an emphasis on the
effects of magnetic fields.  While impressive progress has been made
in the area of isotropic turbulence, both in the HD and the MHD case,
more complex scenarios remain elusive. The treatment of thermal and
chemical processes is still fairly approximate. Our understanding of
self-gravitating turbulence remains poor, despite many ongoing
efforts. Radiation HD and MHD in combination with
adaptive methods, are just beginning.

There is one aspect that should be highlighted whenever we are talking
about the simulation of turbulence in the interstellar medium. The
huge range of length scales from the galactic disk down to
proto-stellar cores does not allow for coverage by a single numerical
simulation. In general, there are two numerical cutoffs, one toward larger scales,
the other toward smaller scales. The former poses the question of
initial as well as boundary conditions and whether simple models such
as periodic boxes and random forcing can account for those in a
self-consistent fashion. The small-scale cutoff would, in general,
necessitate closures. These are considered as essential in atmospheric
sciences but have met very little attention in astrophysics.

As regards the large scales, it certainly will not be feasible in the
near future to run a simulation of a disk galaxy or even a
piece of a galactic disks including all processes of turbulence
production, the different phases of the interstellar medium, the
formation of molecular clouds and the collapse of gas giving rise to
the birth of stars. If energy is injected via radom forcing
in a simulation of turbulence over some subrange of scales in the ISM,
one faces the question of what the influence of the forcing might be. 
That the outcome can vary depending on the applied forcing
is illustrated by a simple numerical experiment. Figure~\ref{fg:forcing} shows
projected mass densities obtained from $128^{3}$ simulations, in
which purely solenoidal (left) and compressive (right) forcing was
applied. Although there is no fully developed turbulence at such low
resolution, the plots suggest that the flow morphology is genuinely
different, with markedly higher density contrasts and more pronounced
intermittency in the case of compressive forcing. Indeed, this
conclusion is confirmed by high-resolution simulations
\cite{SchmFeder07b}.

For the high-fidelity treatment of small scale effects, AMR emerges as
the most promising tool, simply following the maxim to go to very high
levels of refinement wherever the critical events take place.  This
was most impressively executed by Abel et al.\ \cite{AbelBry02} in AMR
simulations of the formation of the first stars in the
Universe. However, circumstances are not that favourable when it comes
to galactic star formation and it remains to be seen whether AMR by
itself will hold its promises. An alternative approach was suggested
by Niemeyer et al.\ \cite{NieSchm05}. The basic idea is to combine the
techniques of AMR, which have been brought to great success in
astrophysics, and subgrid scale modelling, which is commonly used in
engineering. Furthermore, the incorporation of additional physics
such as thermal processes, complex chemistry as well as radiation
transport in AMR simulations are essential for realistic models of
turbulence in the ISM.

\end{document}